\documentclass[10pt,aps,prd,twocolumn,nofootinbib,noshowkeys,noshowpacs,longbibliography,
superscriptaddress,tightenlines,floatfix]{revtex4-2}
\usepackage{amsmath,amsfonts,amsthm,amssymb}
\usepackage{graphics,graphicx}
\usepackage{color}
\definecolor{darkblue}{RGB}{0,0,196}
\definecolor{darkgreen}{RGB}{0,120,0}

\newcommand{\nn}{\nonumber}

\def\beq{\begin{equation}}
\def\eeq{\end{equation}}
\def\st{\begin{equation}}
\def\stp{\end{equation}}
\def\ba{\begin{eqnarray}}
\def\ea{\end{eqnarray}}

\usepackage[colorlinks=true,linktocpage=true,linkcolor=darkblue,citecolor=red,urlcolor=darkblue]{hyperref}

\newcommand{\bra}[1]{\langle #1|}
\newcommand{\ket}[1]{|#1\rangle}

\newcommand{\di}{{\rm d}}

\def\spt{{\cal S}}
\def\wT{{\widehat T}}
\def\wj{{\widehat j}}
\def\wJ{{\widehat J}}

\def\wP{{\widehat P}}

\def\wQ{{\widehat Q}}
\def\wspt{{\widehat{\cal S}}}

\def\wrho{{\widehat{\rho}}}
\def\wrhol{{\widehat{\rho}_{\rm LE}}}

\newcommand{\Tr}{{\rm Tr}}  
  
\newcommand{\e}{{\rm e}}

\newcommand{\be}{\begin{equation}}
\newcommand{\ee}{\end{equation}}                                                                               
\def\bea{\begin{eqnarray}}
\def\eea{\end{eqnarray}}

\begin{document}
%%%%%%%%%%%%%%%%%%%%%
\preprint{}
 
    \title{On the local thermodynamic relations in relativistic spin hydrodynamics}
    % %
    \author{Francesco Becattini}
    \email{becattini@fi.infn.it}
    \affiliation{Department of Physics, University of Florence and INFN Via G. Sansone 1, I-50019, Sesto F.no (Firenze), Italy}
    \author{Rajeev Singh}
    \email{rajeev.singh@e-uvt.ro}
    \affiliation{Department of Physics, West University of Timisoara, Bulevardul Vasile P\^arvan 4, Timisoara 300223, Romania}
	\date{\today} 
	\bigskip
%%%%%
\begin{abstract}
We demonstrate, by providing two specific examples, that the local differential thermodynamic relations 
used as educated guesses in relativistic hydrodynamics with spin, do not hold even at global thermodynamic 
equilibrium. We show, by using a rigorous quantum statistical method, that for massless free fermions 
and massive free fermions with rotation and acceleration at global thermodynamic equilibrium, the derivative 
of the pressure function with respect to the spin potential differs from the spin density and acquires 
a correction of the same order. Such correction cannot be eliminated by any redefinition of the entropy 
current, a so-called entropy-gauge transformation. Therefore, for an accurate determination of the 
constitutive relations in relativistic spin hydrodynamics, the traditional method of assuming differential 
thermodynamic relations is not appropriate.
\end{abstract}
     
\date{\today}
% It is always \today, today, % but any date may be explicitly specified
% \keywords{Relativistic hydrodynamics, causality, stability, field redefinition}
	
\maketitle
% \newpage

%*******************************************************************************
\section{Introduction}
\label{sec:intro}
%********************************************************************************
%
Prompted by the measurements of spin polarization of particles produced in relativistic heavy-ion collisions~\cite{STAR:2017ckg,STAR:2019erd,ALICE:2019onw,ALICE:2021pzu}, there has been a surge 
in the development of relativistic hydrodynamic frameworks that incorporate spin as an additional 
hydrodynamic degree of freedom~\cite{Becattini:2011zz,Florkowski:2017ruc,Florkowski:2018fap,
Montenegro:2017rbu,Hattori:2019lfp, Florkowski:2019qdp,Singh:2020rht, Fukushima:2020ucl,Hongo:2021ona,Gallegos:2021bzp,Florkowski:2021wvk,Gallegos:2022jow,Peng:2021ago,Singh:2025hnb,
Cao:2022aku,Weickgenannt:2022zxs,Weickgenannt:2022qvh,Biswas:2023qsw,Buzzegoli:2026oxc}. 

Relativistic spin hydrodynamics, in essence, includes the conservation of the angular momentum-boost
tensor besides that of energy-momentum. The operator angular momentum-boost density reads:
\bea
 \widehat{\cal J}^{\lambda\mu\nu}= x^\mu \wT^{\lambda\nu} - x^\nu \wT^{\lambda \mu}
 + \wspt^{\lambda\mu\nu} \equiv \widehat{\cal L}^{\lambda\mu\nu} + \wspt^{\lambda\mu\nu} \; , 
\eea
where $\widehat{\cal L}^{\lambda\mu\nu}$ is the so-called orbital part and $\wspt^{\lambda\mu\nu}$ 
is the spin tensor. Due to the conservation of $\widehat{\cal J}^{\lambda\mu\nu}$, the divergence 
of the spin tensor equals the anti-symmetric part of the energy-momentum tensor
\be
\partial_{\lambda}\wspt^{\lambda\mu\nu}=\wT^{\nu\mu}-\wT^{\mu\nu}\,.
\ee
As it is well known, the energy-momentum and spin tensors are not uniquely defined and one may obtain 
different pairs of those tensors through a so-called pseudo-gauge transformation~\cite{Halbwachs,Hehl:1976vr,Becattini:2011ev}
\ba
\hat{T}^\prime{^{\mu\nu}} &=& \hat{T}^{\mu\nu} + \frac{1}{2} \partial_\lambda (\hat{\Pi}^{\lambda,\mu\nu}+\hat{\Pi}^{\nu,\mu\lambda}+\hat{\Pi}^{\mu,\nu\lambda})\,,\\
\hat{\mathcal{S}}^\prime{^{\lambda,\mu\nu}} &=& \hat{\mathcal{S}}^{\lambda,\mu\nu} - 
\hat{\Pi}^{\lambda,\mu\nu} + \partial_\rho \hat{\Theta}^{\mu\nu,\lambda\rho}\,,
\label{eq:pseudogauge_transformations}
\ea
where $\hat{\Pi}$ and $\hat{\Theta}$ are known as super-potentials satisfying, $\hat{\Pi}^{\lambda,\mu\nu} = -\hat{\Pi}^{\lambda,\nu\mu}$ and $\hat{\Theta}^{\mu\nu,\lambda\rho} = - \hat{\Theta}^{\nu\mu,\lambda\rho} = - \hat{\Theta}^{\mu\nu,\rho\lambda}$, respectively. The new pair of energy-momentum and spin tensors, 
however, satisfy the same dynamical equations and do not change the integrated conserved charges.
However, both the entropy and entropy production rate are apparently dependent on the choice
of the pseudo-gauge \cite{Becattini:2023ouz}, thus making the study of the entropy current in
spin hydrodynamics an important subject.

Requiring the local entropy production rate to be positive definite, there have been numerous 
derivations of the constitutive relations of the spin tensor and the energy-momentum tensor as a 
function of the gradients of the thermo-hydrodynamic fields~\cite{Hattori:2019lfp,Fukushima:2020ucl,She:2021lhe,Hongo:2021ona,Daher:2022xon,
Gallegos:2022jow,Biswas:2023qsw}.
In those formulations, the entropy current is obtained by assuming the validity of the 
thermodynamic relations:
\ba
Ts  &=& \rho + p - \mu n - \frac{1}{2} \omega_{\mu\nu} S^{\mu\nu}\,\label{tradtherm}  \,,\\
\di p &=& s \, \di T + n \, \di \mu +\frac{1}{2} S^{\mu\nu} \di \omega_{\mu\nu} \,,
\label{traddifftherm}
\ea
with $s$, $n$, $\rho$, $p$, and $S^{\mu\nu} \equiv \spt^{\lambda,\mu\nu}u_\lambda$ being the 
entropy density, charge density, energy density, pressure, and spin density, respectively, 
whereas $T$, $\mu$, and $\omega_{\mu\nu}$ are temperature, chemical potential, and spin 
potential, respectively~\footnote{In this work, we will name the quantity $\omega = \Omega \, T$
as spin potential, whereas $\Omega$ will be referred to as {\em reduced} spin potential. Note that
in previous literature, $\Omega$ is referred to as either {\it spin potential} or {\it spin chemical 
potential}, but these terms do not accord well to the usual naming of $\mu$ as chemical potential. 
The ratio $\mu/T=\zeta$ will be referred to as {\em reduced} chemical potential.}.

However, for a given choice of the pseudo-gauge, it was argued in ref.~\cite{Becattini:2023ouz} 
that the differential relation \eqref{traddifftherm} does not generally apply, so that it is not suitable 
to derive the entropy production rate, whence the constitutive equations of relativistic spin 
hydrodynamics at the needed level of precision. The inadequacy of the equation \eqref{traddifftherm} was 
also the subject of a recent study \cite{Florkowski:2024bfw,Drogosz:2024gzv} where the authors proposed 
to replace it with a differential relation for vector currents, which we will not delve into in this work.

In ref.~\cite{Becattini:2023ouz} it was shown that while it is possible to define pressure 
such that the \eqref{tradtherm} holds, as well as $\partial p/\partial T|_{\mu,\omega}=s$, it 
is not guaranteed that: 
\bea
\frac{\partial p}{\partial \mu}\Big|_{T,\omega} &=& n\,, \label{nderiv}\\ 
\frac{\partial p}{\partial \omega_{\lambda\nu}}\Big|_{T,\mu} &=& S^{\lambda\nu}\,.\label{omegaderiv}
\eea
The argument is based on a first-principle derivation of
the entropy current within a quantum-relativistic statistical mechanics framework proposed in 
ref.~\cite{Becattini:2019poj} and this method was further applied to derive the entropy 
production rate and the constitutive equations in relativistic spin hydrodynamics~\cite{Daher:2025pfq}.
It can be figured out that a simple addition of the term $-\frac{1}{2}\omega_{\mu\nu} S^{\mu\nu}$ 
in the usual thermodynamic relation~\eqref{tradtherm} does not necessarily imply the validity of 
the eq.~\eqref{omegaderiv} with a simple argument, which applies at global equilibrium, where 
the reduced spin potential is equal to the thermal vorticity tensor $\varpi$, as we will see. 
The spin density $S^{\mu\nu}$ is, at the leading order, first order in vorticity, so in 
order to test the \eqref{traddifftherm} one needs to know all thermodynamic functions 
$s$, $\rho$, $p$, $n$, and $S^{\mu\nu}$ at least up to second order in vorticity. Even if we {\em define} 
one of them, e.g. $p$ or $s$, by using the \eqref{tradtherm}, for the \eqref{omegaderiv} to hold 
we would need that the independently calculated thermodynamic functions fulfill a special 
constraint. For instance, in case of using $\eqref{tradtherm}$ to define $p$:
$$
   T \frac{\partial s}{\partial \omega_{\mu\nu}}\Big|_{T,\mu} - 
   \frac{\partial \rho}{\partial \omega_{\mu\nu}}\Big|_{T,\mu} + 
   \mu \frac{\partial n}{\partial \omega_{\mu\nu}}\Big|_{T,\mu} 
   +\frac{1}{2} \omega_{\rho\sigma} \frac{\partial S^{\rho\sigma}}
   {\partial \omega_{\mu\nu}}\Big|_{T,\mu}= 0,
$$
which is not guaranteed. The same conclusion applies if $s$ was defined through the 
\eqref{tradtherm} and $p$ was known. 

In this paper, we will check the validity of the equation \eqref{omegaderiv} by calculating 
the thermodynamic functions at the second order in vorticity within a quantum statistical 
framework and obtain two specific instances of the breaking of the relation \eqref{omegaderiv}:
for massless free fermions at global thermodynamic equilibrium with rotation 
and acceleration in the canonical pseudo-gauge, for which the exact solution is known; and 
for a system of massive free fermions at global thermodynamic equilibrium with rotation and 
acceleration, again, in the canonical pseudo-gauge, for which we 
know the expression of the thermodynamic functions up to second order in vorticity. 
We will further show that, in the examined cases, no redefinition of the entropy current 
can restore the differential relations \eqref{traddifftherm} and so we will conclude that
at least one of them is incorrect, with all due implications.

%***********************************************************************************
\section{Local equilibrium density operator and the entropy current}
\label{sec:local_eq_den_op}
%***********************************************************************************

To make this article self-contained, we will first briefly review the derivation of the entropy 
current presented in~\cite{Becattini:2023ouz}. 

For a relativistic fluid, by applying the quantum-statistical method, one obtains the local 
equilibrium density operator, $\wrho_{\rm LE}$, by maximizing the entropy:
\begin{equation}
    S = - \Tr \left(\wrho \, \log \wrho \right)\,,
\end{equation}
over some space-like hypersurface $\Sigma$ with normal unit vector $n$ constraining the average values 
of the energy density, momentum density, charge density, and spin density to be the same as their 
original values:
\ba\label{constraints}
 n_\mu T^{\mu\nu} = n_\mu \Tr \left(\wrho\, \wT^{\mu\nu}\right)\,, &\qquad& n_\mu j^{\mu} = 
 n_\mu \Tr \left(\wrho\, \wj^\mu\right) \,,\nonumber\\  n_\mu \spt^{\mu\lambda\nu} &=& 
 n_\mu \Tr \left(\wrho \, \wspt^{\mu\lambda\nu}\right)\,,
\ea
where $\wT^{\mu\nu}$,  $\wspt^{\mu\lambda\nu}$ are the energy-momentum and spin tensor 
operators and $\wj^\mu$ is the vector-current operator. The density operator which solves 
the problem is the local equilibrium density operator:
\begin{align}
\label{densityoperator}
    \wrho_{\rm LE}=\frac{1}{Z_{\rm LE}}\exp\left[-\hat{\Upsilon}\right]\,,
\end{align}
with:
\begin{align}
\label{eq:Upsilonoperator}
    \hat{\Upsilon}=\int_{\Sigma} \di\Sigma_{\mu}\,\left(\wT^{\mu\nu}\beta_{\nu}
    -\zeta \wj^\mu - \frac{1}{2}\Omega_{\lambda\nu}\wspt^{\mu\lambda\nu}\right)\,.
\end{align}
In the equation \eqref{eq:Upsilonoperator} $\beta_{\nu}$, $\zeta$ and $\Omega_{\lambda\nu}$ 
are the Lagrange multiplier fields defined, respectively, as the four-temperature vector, 
the reduced chemical potential and the reduced spin potential:
\be\label{lagrange}
 \beta^\mu = \frac{u^\mu}{T}\,, \qquad \zeta = \frac{\mu}{T}\,, \qquad 
 \Omega^{\lambda\nu} = \frac{\omega^{\lambda\nu}}{T}\,,
\ee
so that $T=1/\sqrt{\beta^2}$ is the proper temperature.
At the global thermodynamic equilibrium, $\beta_\mu$ is a Killing vector, $\Omega^{\mu\nu}$
is constant and equals thermal vorticity $\varpi^{\mu\nu}$ and $\zeta$ is constant:
\ba\label{equilibrium}
 \beta_\mu &=& b_\mu + \varpi_{\mu\nu} x^\nu\,, \quad {\rm with} \quad b_\mu = {\rm const.}\,, \\
   \Omega_{\mu\nu} &=& \varpi_{\mu\nu} = -\frac{1}{2}\left( \partial_\mu \beta_\nu - 
   \partial_\nu\beta_\mu \right) = {\rm const.}\,,\\ 
   \zeta &=& {\rm const.}\,,
\ea
and in this case the local equilibrium density operator becomes:
\be\label{rhoglobal}
 \wrho_{\rm GE}=\frac{1}{Z_{\rm GE}} \exp\left[- b \cdot \wP +\frac{1}{2} \varpi : \wJ + \zeta \wQ
 \right]\,,
\ee
where the subscript `GE' stands for global equilibrium and $\wP$, $\wJ$, and $\wQ$ are the four-momentum operator, boost-angular momentum operator, and conserved charge, respectively.
There are other useful relations at global equilibrium involving the acceleration field 
$a^\mu$ and vorticity field $\omega^\mu$:
\bea\label{accvort}
  \alpha^\mu &\equiv& \varpi^{\mu\nu} u_\nu = \frac{1}{T} u^\nu \partial_\nu u^\mu = 
  \frac{a^\mu}{T}\,, \\ \nonumber
  w^\mu &\equiv& -\frac{1}{2} \epsilon^{\mu\nu\rho\sigma} \varpi_{\nu\rho} u_\sigma
   = -\frac{1}{2T} \epsilon^{\mu\nu\rho\sigma} \partial_\rho u_\nu u_\sigma = \frac{\omega^\mu}{T}\,,
\eea
hence, since at global equilibrium $\varpi_{\mu\nu} = \Omega_{\mu\nu}=\omega_{\mu\nu}/T$, 
we have a very useful formula for the spin potential at global equilibrium:
\be
\omega_{\mu\nu} = a_\mu u_\nu - a_\nu u_\mu + \epsilon_{\mu\nu\rho\sigma} 
\omega^\rho u^\sigma \,.\label{eq:relations_identities}
\ee
It is convenient to define a normalized mean value of operators at local equilibrium:
\be\label{leqmean}
 \mathcal{X}_{\rm LE} \equiv \Tr \left(\wrho_{\rm LE} \widehat {\mathcal{X}}\right) - 
 \bra{0} \widehat {\mathcal{X}} \ket{0}\,,
\ee
where the state $\ket{0}$ in Eq. \eqref{leqmean} is the (supposedly non-degenerate) lowest 
lying state of the operator $\widehat{\Upsilon}$ in the equation \eqref{eq:Upsilonoperator}.
Note that this state does not necessarily coincide with the Minkowski vacuum $\ket{0_M}$;
for instance, if:
$$
 \widehat{\Upsilon} = \frac{1}{T_0} \left( \widehat{H} -\mu \widehat{Q} \right)\,,
$$
for the free Dirac field, the state $\ket{0}$ is the Fermi gas at zero temperature. Indeed,
if we take the limit $T \to 0$ in the equation \eqref{densityoperator} we should recover 
the state $\ket{0}$:
\be\label{rholelimit}
 \lim_{T \to 0} \wrhol = \ket{0}\bra{0}\,,
\ee
implying that the local equilibrium values defined above must fulfill the following
limit:
\be\label{limtzero}
 \lim_{T \to 0} \mathcal{X}_{\rm LE} = 0\,.
\ee
The first step in order to derive the entropy current is to prove that the logarithm of the 
partition function is extensive, meaning it can be written as an integral of some four-vector 
field $\phi^\mu$ over the local equilibrium hypersurface $\Sigma$ as:
\ba
\log Z_{\rm LE} = \int_\Sigma \di \Sigma_\mu \; \phi^\mu 
- \bra{0} \widehat \Upsilon \ket{0}\,.
\label{extensive1}
\ea
The field $\phi^\mu$ is defined as the thermodynamic potential current \cite{Becattini:2012tc} and
can be calculated as follows \cite{Becattini:2019poj,Becattini:2023ouz}:
\ba\label{eq:thermodynamic_potential_current}
 &&\phi^{\mu}(x) =\int_{0}^{T(x)} \frac{\di T'}{T^{\prime 2}} \; \Big( u_{\nu}(x) 
  T^{\mu\nu}_{\rm LE}(x)[T',\mu,\omega] \label{phit}\\
   &&-\mu(x) j^\mu_{\rm LE}(x)[T',\mu,\omega] - \frac{1}{2} \omega_{\lambda\nu}(x) 
  \spt^{\mu\lambda\nu}_{\rm LE}(x)[T',\mu,\omega] \Big) \,.
  \nonumber
\ea
In the above equation, the square brackets denote a functional dependence on the thermodynamic
fields $T'$, $\mu$, and $\omega$. We are now in a position to obtain the entropy current. Starting with 
Eq.~\eqref{densityoperator} and using Eqs.~\eqref{leqmean} and \eqref{extensive1} we have:
\ba
S &=& -\Tr \left(\wrho_{\rm LE} \log \wrho_{\rm LE}\right) \,,\nonumber\\
&=& \log Z_{\rm LE} +  \int_\Sigma \di \Sigma_\mu \; 
\Big( \Tr \left(\wrhol \wT^{\mu\nu}\right) \beta_{\nu} -\zeta \Tr \left( \wrhol \wj^\mu \right)\nonumber \\
&&-
  \frac{1}{2}\Omega_{\lambda\nu}\Tr \left( \wrhol \wspt^{\mu\lambda\nu}\right) \Big)\, ,\nonumber \\
  &=& \int_\Sigma \di \Sigma_\mu \;\left(  \phi^\mu + T^{\mu\nu}_{\rm LE}\beta_{\nu} 
  -\zeta j^\mu_{\rm LE} -\frac{1}{2} \Omega_{\lambda\nu}\spt^{\mu\lambda\nu}_{\rm LE} 
  \right)\,, \nonumber\\
  &=& \int_\Sigma \di \Sigma_\mu \;s^\mu\,,
  \label{totalent}
\ea
whence we can define the entropy current as:
\be\label{def1}
 s^\mu = \phi^\mu + T^{\mu\nu}_{\rm LE}\beta_{\nu} -\zeta j^\mu_{\rm LE} -
  \frac{1}{2}\Omega_{\lambda\nu}\spt^{\mu\lambda\nu}_{\rm LE}\,.
\ee
Because of the constraints \eqref{constraints}, the entropy current can be defined as
well as:
\be\label{def2}
 s^\mu = \phi^\mu + T^{\mu\nu} \beta_{\nu} -\zeta j^\mu - \frac{1}{2}\Omega_{\lambda\nu} 
 \spt^{\mu\lambda\nu}\,,
\ee
where $\phi^\mu$ is then written as:
\be\label{phi2}
 \phi^{\mu}=\int_{0}^{T} \frac{\di T'}{T^{\prime 2}} \; \left( T^{\mu\nu}[T'] u_{\nu}
 -\mu j^\mu[T'] - \frac{1}{2} \omega_{\lambda\nu} \spt^{\mu\lambda\nu}[T']
 \right) \, .
\ee
The form \eqref{def2} of the entropy current is to be preferred in an out-of-equilibrium
situation with respect to \eqref{def1} \cite{Becattini:2023ouz}. Nevertheless, at global 
thermodynamic equilibrium
the two definitions coincide and there is no ambiguity in its identification. It is worth
pointing out that the entropy current, by virtue of the definition \eqref{def1}, vanishes
in the limit $T \to 0$:
$$
\lim_{T \to 0} s^\mu = 0\,,
$$
as it should because the entropy is zero when $\wrhol$ becomes the pure state $\ket{0}$,
see equation \eqref{rholelimit}.
This condition entails that, as also implied by equation \eqref{constraints}, the 
mean values of the conserved currents $T^{\mu\nu}$, $j^\mu$ etc. vanish in the same limit. In
other words, it is understood that in the formulae \eqref{def2}, \eqref{phi2} they are 
normalized so as to make their value vanishing at zero temperature. Therefore, they
do not necessarily vanish in the Minkowski vacuum, as implied by the above 
discussion.

%***********************************************************************************************
\section{Local thermodynamic relations for a free Dirac field at global equilibrium}
\label{thermodynamic_relations}
%***********************************************************************************************

To obtain the local thermodynamic relations, one requires to contract the entropy current~\eqref{def2} 
with the fluid four-velocity, $u^\mu = \beta^\mu/\sqrt{\beta^2} = T \beta^\mu$ (normalized as $u \cdot u = 1$) \footnote{Again, we stress
that no ambiguity exists in the definition of a four-velocity at global thermodynamic equilibrium, 
where $\beta$ is a Killing vector.}
\ba\label{ltr1}
  s \equiv s^\mu u_\mu &=& \phi^\mu u_\mu + T^{\mu\nu} \beta_{\nu}u_\mu -\zeta j^\mu u_\mu- \frac{1}{2}u_\mu\Omega_{\lambda\nu} \spt^{\mu\lambda\nu} \nonumber\\
  &=& \phi \cdot u + \frac{1}{T} \rho - \frac{\mu}{T} n - \frac{1}{2} \frac{\omega_{\lambda\nu}}{T}
  S^{\lambda\nu}\,,
\ea
where $\rho=u_\mu u_\nu T^{\mu\nu}$ is the energy density, $n=u_\mu j^\mu$ is the charge density, and 
$S^{\lambda\nu} = u_\mu \spt^{\mu\lambda\nu}$ is the spin density. Thus, {\em defining} the pressure 
$p$ as $T \phi \cdot u$, we have
\be\label{localtr}
T s = p +  \rho - \mu n - \frac{1}{2} \omega_{\lambda\nu} S^{\lambda\nu}\,,
\ee
which is the first thermodynamic relation in eq.~\eqref{tradtherm}. It should be emphasized that
the hereby defined pressure does not coincide in general with the average diagonal stress in the
energy-momentum tensor. Particularly, it does not need to fulfill the equation $p=\rho/3$ for a
massless Dirac field.

Following this definition and using the eq.~\eqref{phi2} we have:
\begin{eqnarray}\label{pressure}
    p &=& T \phi^\mu u_\mu \,,\\
    &=& T\int^{T}_0 \frac{{\rm d}T^\prime}{{T^\prime}^2} \Big(T^{\mu\nu}[T^\prime] 
    u_\nu u_\mu - \mu u_\mu j^\mu [T^\prime] - \nonumber \\
    &-& \frac{1}{2} \omega_{\lambda\nu} u_\mu \mathcal{S}^{\mu\lambda\nu}[T^\prime] \Big)\,, \nonumber \\
    &=& T\int^{T}_0 \frac{{\rm d}T^\prime}{{T^\prime}^2} \left(\rho[T^\prime] 
     - \mu n[T^\prime] - \frac{1}{2} \omega_{\lambda\nu} S^{\lambda\nu}[T^\prime]\right)\,. \nonumber
    \label{eq:pressure_function}
\end{eqnarray}
While the above equation, alongside with the \eqref{localtr}, implies:
\be\label{dpdt}
 \frac{\partial p}{\partial T}\Big|_{\mu,\omega} = s\,,
\ee
the equations \eqref{nderiv} and \eqref{omegaderiv} cannot be derived from the equation
\eqref{pressure}. In fact, we will show in the following that for a system of massless 
free fermions and massive free fermions at global thermodynamic equilibrium with rotation 
and acceleration, the differential relation \eqref{omegaderiv} does not hold.

%-----------------------------------------------------------------------------
\subsection{Massless Dirac fermions}
%-----------------------------------------------------------------------------

The exact mean values of the currents normalized so as to vanish in the Minkowski vacuum 
$\ket{0_M}$, at global thermodynamic equilibrium with the density operator \eqref{rhoglobal} have been 
derived in refs.~\cite{Palermo:2021hlf,Palermo:2023ews} for the massless Dirac field. 
In order not to confuse them with the mean values normalized so as to vanish in the lowest lying
state of the operator $\widehat \Upsilon$, that is, in the limit $T \to 0$, we will denote them by
a tilde, hence:
$$
    \tilde {\cal X} = \Tr(\wrho_{\rm GE} \widehat{\cal X}) - 
    \bra{0_M} \widehat{\cal X} \ket{0_M}\,.
$$
What we need is, however:
\begin{align*}
    {\cal X} &= \Tr(\wrho_{\rm GE} \widehat{\cal X}) - 
    \bra{0} \widehat{\cal X} \ket{0} \\
    & = \Tr(\wrho_{\rm GE} \widehat{\cal X}) - 
    \bra{0_M} \widehat{\cal X} \ket{0_M} + \bra{0_M} \widehat{\cal X} \ket{0_M}- 
    \bra{0} \widehat{\cal X} \ket{0} \\
    &= \tilde{\cal X} + \bra{0_M} \widehat{\cal X} \ket{0_M}- 
    \bra{0} \widehat{\cal X} \ket{0}\,.
\end{align*}    
From the above equation, making use of the \eqref{limtzero}, we have:
$$
\lim_{T \to 0} \tilde{\cal X} = -\bra{0_M} \widehat{\cal X} \ket{0_M} + 
\bra{0} \widehat{\cal X} \ket{0}\,,
$$
which leads to the useful identity:
\be\label{diffcurre}
   {\cal X} = \tilde{\cal X} - \lim_{T \to 0} \tilde{\cal X}\,,
\ee
providing us with a straightforward method to calculate the mean values appearing
in the entropy equation.
For instance, the vector and axial current normalized so as to vanish in the
Minkowski vacuum read~\cite{Ambrus:2014uqa,Ambrus:2019cvr,Ambrus:2019ayb,Palermo:2021hlf,Palermo:2023ews}:
\begin{eqnarray}
    \tilde j^\mu &=& \frac{\zeta\,T}{\pi^2}\left(\frac{T^2\left(\pi^2 + \zeta^2\right)}{3 } - 
    \frac{a^2 }{4 } + \frac{\omega^2}{4 }\right) u^\mu- \frac{\zeta\, l^\mu\,T^3}{6\pi^2}\,, \nonumber \\
    \tilde j^\mu_A &=& \frac{T^2}{\pi^2}\left( \frac{\pi^2}{6} + \frac{\zeta^2}{2} - 
    \frac{\omega^2 }{24\,  T^2} - \frac{a^2 }{8\, T^2}\right) \omega^\mu\,,
    \label{eq:axial_current}
\end{eqnarray}
where $a^\mu$ and $\omega^\mu$ have been defined in the
eq. \eqref{eq:relations_identities} and:
$$
 l^\mu = \epsilon^{\mu\nu\rho\sigma} w_\nu \alpha_\rho u_\sigma \,.
$$
Applying equation \eqref{diffcurre} to subtract the zero temperature limit, we get 
the following simple forms for the vector and axial currents to be used in the entropy 
current calculation:
\begin{eqnarray}\label{axial2}
  j^\mu &=& \frac{1}{3} \mu T^2 u^\mu\,, \nonumber \\
  j^\mu_A &=& \frac{1}{6} T^2 \omega^\mu\,.
\end{eqnarray}
In refs.~\cite{Palermo:2021hlf,Palermo:2023ews} only the exact expression of the symmetric
Belinfante energy-momentum tensor was derived. However, since we need to tackle thermodynamic 
relations with a spin tensor, we have to select a pseudo-gauge where the spin tensor 
does not vanish. For the free Dirac field, the simplest pseudo-gauge with a non-trivial 
spin tensor is the canonical one, which is related to the Belinfante pseudo-gauge by the 
following equation:
\be
 T^{\mu\nu}_{\rm Can} =  T^{\mu\nu}_{\rm B} - 
\frac{1}{2} \partial_\lambda \mathcal{S}^{\lambda\mu\nu}_{\rm Can}\,.
\label{eq:canonicalTmunu}
\ee
where the subscript `Can' represents canonical currents and `B' stands for Belinfante. 
The mean value of the canonical spin tensor is related to the axial current:
\be\label{eq:spin_tensor_canonical}
\mathcal{S}^{\lambda\mu\nu}_{\rm Can} = -\frac{1}{2}\epsilon^{\lambda\mu\nu\rho} 
j_{A \rho} = -\frac{1}{12}\epsilon^{\lambda\mu\nu\rho} T^2 \omega_\rho\,,
\ee
where we used the axial current~\eqref{axial2}. Now, contracting Eq.~\eqref{eq:canonicalTmunu} 
with $u_\mu u_\nu$ and using the Belinfante energy-momentum tensor derived in 
\cite{Palermo:2023ews} (see Appendix \ref{app:A}) we obtain:
\begin{align}\label{contr1}
      & \rho = u_\mu u_\nu T^{\mu\nu}_{\rm Can} = u_\mu u_\nu T^{\mu\nu}_{\rm B} - 
   \frac{1}{2} u_\mu u_\nu \partial_\lambda \mathcal{S}^{\lambda\mu\nu}_{\rm Can} \\
    &= u_\mu u_\nu T^{\mu\nu}_{\rm B} = \frac{7\pi^2 T^4}{60}  + \frac{\mu^2 T^2}{2}  
    - \frac{a^2 T^2}{24}  - \frac{\omega^2 T^2}{8}  ,\nonumber
\end{align}
where we took advantage of the antisymmetry of the indices in \eqref{eq:spin_tensor_canonical}.
By using the equations \eqref{axial2} and \eqref{eq:spin_tensor_canonical} we can obtain the
contractions:
\begin{align} \label{contr2}
  n &= u_\mu j^\mu = \frac{1}{3} \mu T^2 \,, \\
  \omega_{\mu\nu} S^{\mu\nu} &= - \frac{T^2}{12}\epsilon^{\lambda\mu\nu\rho} 
  \omega_\rho \omega_{\mu\nu} u_\lambda\,,\nonumber \\
  &= - \frac{T^2}{12}\epsilon^{\lambda\mu\nu\rho} 
  \epsilon_{\mu\nu\alpha\beta}\omega_\rho \omega^\alpha u^\beta u_\lambda
  = -\frac{T^2}{6} \omega^2\,.
  \label{contr3}
\end{align}
With the equations \eqref{contr1}, \eqref{contr2} and \eqref{contr3} we are now in a 
position to determine the pressure by integrating the right-hand side of equation \eqref{pressure}. 
Splitting the pressure of equation \eqref{pressure} into three terms:
\be\label{psplit}
p = p_1 - p_2 - \frac{1}{2} p_3\,,
\ee
we get:
\begin{eqnarray}\label{press3}
p_1 &=& T\int^{T}_0 \frac{{\rm d}T^\prime}{{T^\prime}^2} \rho[T'] \\
&=& \frac{7\pi^2 }{180}T^4 + \frac{1}{2} \mu^2 T^2  - \frac{{a}^2}{24} T^2 
- \frac{{\omega}^2}{8}T^2\,, \nonumber \\
p_2 &=& T\int^{T}_0 \frac{{\rm d}T^\prime}{{T^\prime}^2} \mu n [T^\prime] = 
\frac{1}{3} \mu^2 T^2\,, \nonumber \\
p_3 &=& T\int^{T}_0 \frac{{\rm d}T^\prime}{{T^\prime}^2} \omega_{\mu\nu} S^{\mu\nu} [T^\prime] 
= -\frac{1}{6}\omega^2 T^2\, \nonumber .
\end{eqnarray}
By using the equations \eqref{press3}, the equation \eqref{psplit} yields the pressure:
\begin{eqnarray}\label{eq:pressure_function1}
    p = \frac{7\pi^2}{180} T^4 + \frac{1}{6} \mu^2 T^2  - 
    \left({a}^2+{\omega}^2\right)\frac{T^2}{24}  \,.
\end{eqnarray}
While the entropy density can be obtained from the equation \eqref{localtr} by using the 
\eqref{contr1} and \eqref{contr2}:
$$
 s = \frac{7\pi^2}{45} T^3 + \frac{1}{3} \mu^2 T - \left({a}^2  + {\omega}^2\right)\frac{T}{12}\,.
$$
Now, from \eqref{eq:pressure_function1}
\begin{eqnarray}
    \frac{\partial p}{\partial T}\Big|_{\mu,\omega} = \frac{7\pi^2 {T}^3}{45} + 
    \frac{1}{3} \mu^2 T  - \left({a}^2  + {\omega}^2\right)\frac{T}{12} \equiv s \,,
    \label{eq:s}
\end{eqnarray}
as expected, and:
\begin{eqnarray} \label{eq:n}
 \frac{\partial p}{\partial \mu}\Big|_{T,\omega} = \frac{1}{3} \mu T^2 \equiv n\,,
\end{eqnarray}
which is also in agreement with the differential relation \eqref{tradtherm}. Yet:
\begin{eqnarray}\label{eq:spin}
\frac{\partial p}{\partial \omega_{\lambda\nu}}\Big|_{T,\mu} &=& \frac{\partial p}{\partial {a}^2} 
\frac{\partial {a}^2}{\partial \omega_{\lambda\nu}} + \frac{\partial p}{\partial {\omega}^2} 
\frac{\partial{\omega}^2}{\partial \omega_{\lambda\nu}}\,,\\
&=& -\frac{T^2}{12}\left( a_\rho \frac{\partial a^\rho}{\partial \omega_{\lambda\nu}} + 
\omega_\rho \frac{\partial \omega^\rho}{\partial \omega_{\lambda\nu}}\right)\,,\nonumber\\
&=&-\frac{T^2}{12}\left( {a}^\rho u^\sigma \delta^{\lambda\nu}_{\rho\sigma} - 
\frac{1}{2}\epsilon^{\rho\alpha\beta\gamma}{\omega}_\rho u_\gamma \delta^{\lambda\nu}_{\alpha\beta}\right)\,,\nonumber\\
 &=&\frac{T^2}{12}\left({a}^\nu u^\lambda-{a}^\lambda u^\nu + 
 \epsilon^{\lambda\nu\rho\gamma}\omega_\rho u_\gamma \right)\,,\nonumber\\
&=&\frac{T^2}{12}\left({a}^\nu u^\lambda-{a}^\lambda u^\nu \right) + S^{\lambda\nu} \,,\nonumber
\end{eqnarray}
where we have used the equation \eqref{eq:spin_tensor_canonical} and the definition
of the spin density $S^{\lambda\nu} = u_\mu \spt^{\mu\lambda\nu}_{\rm Can}$. Hence, the derivative 
of the pressure with respect to the spin potential does not match the spin density and
the relation \eqref{omegaderiv} does not apply.

%-----------------------------------------------------------------------------------------
\subsection{Massive Dirac fermions}
%-----------------------------------------------------------------------------------------

A similar calculation can be carried out for massive fermions at second order of
the expansion in thermal vorticity at global thermodynamic equilibrium. The general expression, 
up to that order, of the Belinfante energy-momentum tensor, the vector and axial currents normalized
with Minkowski vacuum subtraction was obtained in ref.~\cite{Buzzegoli:2017cqy}:
\begin{eqnarray}
    \tilde T^{\mu\nu}_{\rm B}&=& \left(\bar\rho - \frac{a^2  U_\alpha}{T^2} - 
    \frac{\omega^2  U_w}{T^2}\right) u^\mu u^\nu + \frac{A}{T^2} a^\mu a^\nu \nn\\
    &-& \left(\bar p - \frac{a^2  D_\alpha}{T^2} - \frac{\omega^2  D_w}{T^2}\right) \Delta^{\mu\nu}+ 
    \frac{W}{T^2} \omega^\mu \omega^\nu \nonumber\\
    &+&  G \left(u^\mu l^\nu + u^\nu l^\mu\right)\,, \label{eq:Tmunu_DiracM}\\
    \tilde j^\mu &=& \bar n \,u^\mu + \left(\frac{a^2  N_\alpha}{T^2} + \frac{\omega^2 N_w}{T^2}\right)u^\mu 
    + G_V l^\mu\,, \label{vectorcurrent}\\
    \tilde j^\mu_A &=& \frac{1}{T}\omega^\mu W^A\,.
    \label{axialcurrent}
\end{eqnarray}
The expression of the coefficients $\bar\rho$, $\bar p$, $U_\alpha$, $U_w$, $A$, $D_\alpha$, $D_w$, $W$, 
$G$, $\bar n$, $N_\alpha$, $N_w$, $G_V$, and $W_A$ 
is reported in Appendix \ref{app:B} in the equation \eqref{eq:coefficients_general}. By using 
the equations \eqref{eq:canonicalTmunu} and \eqref{eq:spin_tensor_canonical} we find:
\begin{eqnarray}
    u_\mu u_\nu \tilde T^{\mu\nu}_{\rm Can} &=& \bar \rho - \frac{a^2 U_\alpha}{T^2}-\frac{\omega^2 
     U_w}{T^2} \,,\nonumber\\
    u_\mu \tilde j^\mu &=& \bar n + \frac{1}{T^2} \left( a^2  N_\alpha + \omega^2 N_w\right)\,,\nonumber\\
    u_\mu \omega_{\lambda\nu} \tilde{\mathcal{S}}^{\mu\lambda\nu}_{\rm Can} &=& -\frac{1}{T}\omega^2 W^A\,.
    \label{eq:massive_functions}
\end{eqnarray}
One can work out the coefficients in eq.~\eqref{eq:coefficients_general}, in the 
ultra-relativistic and the non-relativistic limits.

The ultra-relativistic limit $m\rightarrow 0$ is just a control test, because it should
coincide with our previous calculation. Indeed, it can be shown that in this limit the
coefficients in eq. \eqref{eq:coefficients_general} reduce to:
\begin{eqnarray}
\bar \rho &=& \frac{7\pi^2 T^4}{60}+\frac{\mu^2 T^2}{2}+\frac{\mu^4}{4 \pi^2}  ,\quad
U_\alpha = \left(\frac{T^4}{24}+\frac{\mu^2 T^2}{8\pi^2}\right),\nonumber\\
U_w &=& \frac{1}{8}\left(T^4 + \frac{3 \mu^2 T^2}{\pi^2} \right)\,,\quad
  W^A = \frac{T^3}{6}+\frac{T \mu^2}{2 \pi^2}\,,\label{eq:coefficients_mZero_case}\\
  \bar n &=& \frac{T^2 \mu}{3 }\left(1+ \frac{\mu^2}{\pi^2 T^2} \right)\,,\quad
  N_\alpha = N_w = \frac{T^2 \mu}{4\pi^2}\,.\nn
\end{eqnarray}
After subtraction of the zero temperature limit as prescribed, eqs.~\eqref{eq:massive_functions} 
give rise to:
\begin{eqnarray}
u_\nu u_\mu T^{\mu\nu}_{\rm Can}  &=&  \rho = 
\frac{7\pi^2 T^4}{60}+\frac{\mu^2 T^2}{2}- \frac{a^2}{24} T^2-\frac{\omega^2}{8}T^2 ,\nonumber\\
u_\mu j^\mu &=& n = \frac{T^2 \mu}{3 },\nonumber\\
u_\mu \omega_{\lambda\nu} \mathcal{S}^{\mu\lambda\nu}_{\rm Can} &=&  -\omega^2 \frac{T^2}{6}.
\end{eqnarray}
The ensuing pressure function defined in eq.~\eqref{pressure} fully matches the one in eq.~\eqref{eq:pressure_function1}.

After having successfully checked the ultra-relativistic limit, we are now in a position
to study the non-relativistic case $m/T \gg 1$. Setting $x=m/T$, we use the asymptotic 
expansion of the McDonald functions as~\cite{Buzzegoli:2017cqy}
\begin{equation}
    K_\nu(n x) \simeq \e^{-nx}\sqrt{\frac{\pi}{2x}}\left[1+\frac{4 \nu^2-1}{8nx}+\dots\right].
\end{equation}
Thus, retaining only the dominant contribution in the non-relativistic limit, with $\mu < m$
or $\zeta < x$ (see definition \eqref{lagrange}) the coefficients in eq.~\eqref{eq:coefficients_general} 
are well approximated by:
\begin{eqnarray}\label{apprexp}
\bar\rho &=& \frac{m^4}{\sqrt{2}(\pi x)^{3/2}} \e^{-(x-\zeta)} \,,\nonumber\\
U_\alpha &=& \frac{x\,m^4}{24\sqrt{2}(\pi x)^{3/2}} \e^{-(x-\zeta)} \,,\nonumber\\
U_w &=& \frac{m^4}{8\sqrt{2}(\pi x)^{3/2}} \e^{-(x-\zeta)} \,,\nonumber\\
  W^A &=&\frac{m^3}{2\sqrt{2}(\pi x)^{3/2}} \e^{-(x-\zeta)}  \,,\nonumber\\
  \bar n &=& \frac{m^3}{\sqrt{2}(\pi x)^{3/2}} \e^{-(x-\zeta)} \,,\nonumber\\
  N_\alpha &=& \frac{x\,m^3}{24\sqrt{2}(\pi x)^{3/2}} \e^{-(x-\zeta)}\,,\nonumber\\
  N_w &=& \frac{m^3}{8\sqrt{2}(\pi x)^{3/2}} \e^{-(x-\zeta)} \,,
  \label{eq:coefficients_non-rel}
\end{eqnarray}
so that the eqs.~\eqref{eq:massive_functions} become:
\begin{eqnarray}\label{nrellim}
    u_\nu u_\mu \tilde T^{\mu\nu}_{\rm Can}  &=&  \frac{m^4 \e^{-(x-\zeta)}}
    {\sqrt{2}(\pi x)^{3/2}} \left(1- \frac{a^2  x}{24\,T^2}- \frac{\omega^2}{8\,T^2}\right) ,\nonumber\\
    u_\mu \tilde j^\mu &=&   \frac{m^3 \e^{-(x-\zeta)}}{\sqrt{2}(\pi x)^{3/2}}
    \left(1+ \frac{a^2 x}{24\,T^2}+ \frac{\omega^2}{8\,T^2}\right) ,\nonumber\\
    u_\mu \omega_{\lambda\nu} \tilde{\mathcal{S}}^{\mu\lambda\nu}_{\rm Can} &=&  - 
    \frac{m^3\,\omega^2}{2\sqrt{2} \, T (\pi x)^{3/2}} \e^{-(x-\zeta)}.
\end{eqnarray}
The zero temperature limit vanishes for all quantities in \eqref{nrellim}, so that
the thermodynamic functions $\rho$, $n$ and $\omega_{\lambda\nu} S^{\lambda\nu}$ coincide 
with the right-hand side of the above equations.
Therefore, in the non-relativistic limit, the pressure $p$ defined in eq.~\eqref{pressure} 
reads:
\begin{eqnarray}
    p &=& \frac{m \e^{-x} \sqrt{x}}{96 \sqrt{2} \pi ^{3/2} (m-\mu )^3} 
    \big(-m a^2 (\mu  +m) \big(\e^{\zeta} \big(4 (\mu ^2+ m^2)\nonumber\\
    &-&8 \mu  m+6  T(m-\mu)-3 T^2\big)+3  T^{3/2} \sqrt{\pi(m-\mu) } \e^{x}\big)\nonumber\\
    &-&6 T (m-\mu ) \big(\e^{\zeta} \big(\omega ^2 (\mu +m) (2 (m-\mu)-T)\nonumber\\
    &+&4 (\mu-m)   \omega ^2  T-64 T (m-\mu )^3\big)\nonumber\\
    &+& \sqrt{\pi T(m-\mu) } \e^{x} \big(\omega ^2 (\mu +m)+32 (m-\mu )^3\big)\big)\big)  \,.
\end{eqnarray}
We can now evaluate its derivative with respect to $\omega$:
\begin{eqnarray}\label{eq:massive_spin-Dirac}
\frac{\partial p}{\partial \omega_{\lambda\nu}}\Big|_{T,\mu} &=& 
\frac{\partial p}{\partial {a}^2} \frac{\partial{a}^2}{\partial \omega_{\lambda\nu}} + 
\frac{\partial p}{\partial {\omega}^2} \frac{\partial{\omega}^2}{\partial \omega_{\lambda\nu}} \\
&=& 2\frac{\partial p}{\partial {a}^2}\, a \frac{\partial a}{\partial \omega_{\lambda\nu}} 
+ 2\frac{\partial p}{\partial {\omega}^2}\, \omega \frac{\partial \omega}{\partial\omega_{\lambda\nu}} \nonumber\\
&=&2\frac{\partial p}{\partial {a}^2} {a}^\rho u^\sigma \delta^{\lambda\nu}_{\rho\sigma} 
+2\frac{\partial p}{\partial {\omega}^2}  {\omega}_\rho \left(-\frac{1}{2}\epsilon^{\rho\alpha\beta\gamma} u_\gamma\right)\delta^{\lambda\nu}_{\alpha\beta} \nonumber\\
&=&2\frac{\partial p}{\partial {a}^2}\left({a}^\lambda u^\nu-{a}^\nu u^\lambda\right) 
-2\frac{\partial p}{\partial {\omega}^2}  \epsilon^{\rho\lambda\nu\gamma}{\omega}_\rho u_\gamma \,,\nn
\end{eqnarray}
where:
\begin{eqnarray}\label{pderiv}
\frac{\partial p}{\partial {a}^2} &=& -\frac{m^{5/2} (m+\mu)}{96 \pi ^{3/2} \sqrt{2 T} 
(m-\mu )^3} \Big(\e^{(\zeta-x)} \big(4 \left(\mu ^2+m^2\right)\nonumber\\
&-&8 \mu  m+6 T (m-\mu )-3 T^2\big)+3 \sqrt{\pi } T^{3/2} \sqrt{m-\mu }\Big),\nonumber\\
\frac{\partial p}{\partial {\omega}^2}&=&-\frac{m \sqrt{m T}}{16 \sqrt{2} 
\pi ^{3/2} (m-\mu )^2} \big(\e^{(\zeta-x)} \big(2 \left(m^2-\mu ^2\right)\nonumber\\
&-&5 m T+3 \mu  T\big)+(\mu +m) \sqrt{\pi  T (m-\mu )}\big).
\end{eqnarray}
The canonical spin density may be obtained like in the equation \eqref{eq:spin_tensor_canonical}
by using the \eqref{axialcurrent} and the approximated expression in \eqref{apprexp} for
$W_A$:
\bea\label{spindens}
S^{\mu\nu} &=& u_\lambda \mathcal{S}^{\lambda\mu\nu}_{\rm Can} = -\frac{1}{2}\epsilon^{\lambda\mu\nu\rho} 
u_\lambda j_{A \rho} = -\frac{1}{2T}\epsilon^{\lambda\mu\nu\rho} W_A \omega_\rho u_\lambda \nonumber \\
&\simeq&  -\frac{1}{2T}\epsilon^{\lambda\mu\nu\rho} \omega_\rho u_\lambda  
\frac{m^3}{2\sqrt{2}(\pi x)^{3/2}} \e^{-(x-\zeta)} \,.
\eea
Comparing the \eqref{spindens} with the \eqref{eq:massive_spin-Dirac}-\eqref{pderiv} we
can readily realize that:
\begin{eqnarray}
\frac{\partial p}{\partial \omega_{\lambda\nu}}\Big|_{T,\mu}\neq   S^{\lambda\nu}\,.
\label{eq:spin_massive}
\end{eqnarray}
Again, like in the massless case, in the non-relativistic limit the derivative of the pressure
has an additional component which is proportional to $(a^\lambda u^\nu - a^\nu u^\lambda)$ 
of the same order of magnitude as the one proportional to the vorticity. Hence, the same
conclusion holds.

%********************************************************************************************
\section{Entropy-gauge transformations}
\label{sec:egauge}
%********************************************************************************************

There is a certain freedom in defining the entropy current $s^\mu$. All in all, its main purpose 
is to yield the total entropy $S$ by integrating it over a space-like 3D hypersurface:
$$
   S = \int_\Sigma \di \Sigma_\mu \; s^\mu \; .
$$
At global thermodynamic equilibrium, the total entropy $S$ is independent of the hypersurface
$\Sigma$, which requires that the entropy current has zero divergence, $\partial_\mu s^\mu = 0$. 
That being said, the entropy current can be modified by adding the divergence of an anti-symmetric 
field $A^{\lambda\mu}$:
\be\label{engauge}
  s^{\prime\mu} = s^\mu + \partial_\lambda A^{\lambda\mu}\,,
\ee
which leaves its divergence invariant as well as the total entropy provided that suitable 
boundary conditions for the field $A$ are set. The transformation~\eqref{engauge} of the entropy 
current has been named 
{\em entropy-gauge transformation} in ref.~\cite{Becattini:2023ouz}. Indeed, one may wonder 
whether through an entropy-gauge transformation the traditional relations \eqref{tradtherm} 
and \eqref{traddifftherm} can be restored. To begin with, the thermodynamic potential current 
$\phi^\mu$ can be modified likewise so as to maintain the relation \eqref{extensive1} as well
as the \eqref{def2}; in symbols:
\be\label{phigauge}
  \phi^{\prime\mu} = \phi^\mu + \partial_\lambda A^{\lambda\mu}\,,
\ee
With the \eqref{engauge} and \eqref{phigauge} the \eqref{def2} is invariant and can
be rewritten as:
\be
 s^{\prime\mu} = \phi^{\prime\mu} + T^{\mu\nu} \beta_{\nu} -\zeta j^\mu - \frac{1}{2}\Omega_{\lambda\nu} 
 \spt^{\mu\lambda\nu}\,.
\ee
To keep the local thermodynamic relation in the form of \eqref{tradtherm} one must redefine 
the pressure $p'= T u_\mu \phi^{\prime\mu}$. The question is whether with such a new definition
it is possible to implement the \eqref{traddifftherm}, notably:
$$
 \frac{\partial p'}{\partial T}\Big|_{\mu,\omega} = s'\,, \qquad \frac{\partial p'}{\partial \omega_{\lambda\nu}}\Big|_{T,\mu} = S^{\lambda\nu}\,.
$$
Taking into account the equations \eqref{dpdt} and \eqref{engauge}, the first equation on the 
left requires \cite{Becattini:2023ouz}:
\be\label{cond1}
 \frac{\partial}{\partial T} \left( u_\mu \partial_\lambda A^{\lambda\mu} \right) \Big|_{\mu,\omega} \equiv
 \frac{\partial \Gamma}{\partial T}\Big|_{\mu,\omega} = 0 \,,
\ee
while the second:
\be\label{cond2}
  \frac{\partial \Gamma}{\partial \omega_{\lambda\nu}}\Big|_{T,\mu} = S^{\lambda\nu}
  - \frac{\partial p}{\partial \omega_{\lambda\nu}}\Big|_{T,\mu}\,.
\ee
The next question is whether such a function $\Gamma$ exists. By taking the derivative 
of the \eqref{cond2} with respect to $T$ and of the \eqref{cond1} with respect to 
$\omega_{\lambda\nu}$, we derive a necessary condition for its existence, namely:
\be\label{necess}
  \frac{\partial}{\partial T} \left( S^{\lambda\nu}-\frac{\partial p}{\partial \omega_{\lambda\nu}}\Big|_{T,\mu}
  \right) \Big|_{\mu,\omega} = 0\,,
\ee
which is equivalent to the second-order mixed derivative condition implied by the 
\eqref{traddifftherm}:
$$
 \frac{\partial S^{\lambda\nu}}{\partial T} \Big|_{\mu,\omega} = 
 \frac{\partial s}{\partial \omega_{\lambda\nu}}\Big|_{T,\mu}\,.
$$
In the massless fermion case, according to the equation \eqref{eq:spin} we have:
\begin{align*}
S^{\lambda\nu}-\frac{\partial p}{\partial \omega_{\lambda\nu}}\Big|_{T,\mu} &= 
\frac{T^2}{12} (a^\lambda u^\nu - a^\nu u^\lambda)\,, \\  
&= \frac{T^2}{12} (\omega^{\lambda\mu} u_\mu u^\nu - \omega^{\nu\mu} u_\mu u^\lambda)\,,
\end{align*}
which does not manifestly fulfill equation \eqref{necess}. The same conclusion is achieved
for the massive fermion case, by using the equations \eqref{eq:massive_spin-Dirac} and
\eqref{spindens}. Therefore, we conclude that a redefinition of the entropy current cannot restore
the differential thermodynamic relations \eqref{traddifftherm}.

%********************************************************************************************
\section{Conclusions}
\label{sec:conclude}
%********************************************************************************************
%
In conclusion, we have shown, by providing two specific counter-examples, namely a system of 
massless free fermions and massive free fermions with rotation and acceleration at global 
thermodynamic equilibrium in the canonical pseudo-gauge, that the differential thermodynamic relations~\eqref{traddifftherm} often used to obtain the constitutive equations in relativistic 
spin hydrodynamics do not hold. 
Notably, the derivative of the pressure function with respect to spin potential (which reduces 
to vorticity at global equilibrium) features additional contributions of the same order of magnitude 
as the spin density. The differential thermodynamic relations cannot be restored by a suitable 
redefinition of the entropy current (an entropy-gauge transformation).\\
The relations \eqref{tradtherm} and \eqref{traddifftherm} therefore can hardly be used to 
make a precise derivation of the constitutive relations in relativistic spin hydrodynamics; their
use may easily lead to missing some dissipative terms for the spin tensor. 
Indeed, the quantum-statistical method proposed in refs.~\cite{Becattini:2023ouz,Daher:2025pfq} to 
determine the entropy production rate and constitutive equations in relativistic spin hydrodynamics 
led to the identification of several new dissipative coefficients for the spin tensor. 

%********************************************************************************************
\begin{acknowledgments}
R.S. acknowledges the kind hospitality and support of INFN Firenze, Department of Physics and 
Astronomy, University of Florence, and ECT* Trento, where part of this work has been carried out. 
R.S. is also supported partly by a postdoctoral fellowship of West University of Timișoara, 
Romania. F.B. is partly supported by the project PRIN2022 Advanced Probes of the Quark Gluon 
Plasma funded by ”Ministero dell’Università e della Ricerca”, Italy.
We thank the Galileo Galilei Institute for Theoretical Physics for the hospitality 
and the INFN for partial support during the completion of this work. We are grateful to Dirk
Rischke for very useful comments.

\end{acknowledgments}
%***********************************************************************************
\appendix

%***********************************************************************************
\section{Energy-momentum tensor for free massless Dirac field}
\label{app:A}
%***********************************************************************************

For the case of global equilibrium with rotation and acceleration, the exact form of
the Belinfante energy-momentum tensor for massless fermions, normalized with Minkowski 
vacuum subtraction was derived in refs.~\cite{Palermo:2021hlf,Palermo:2023ews}:
\begin{eqnarray}
    \tilde T^{\mu\nu}_{\rm B} &=& \bar \rho \, u^\mu u^\nu - \bar p \Delta^{\mu\nu} + 
    W w^\mu w^\nu + 
    A \alpha^\mu \alpha^\nu + G^l l^\mu l^\nu \nn\\
    &+& G (l^\mu u^\nu + l^\nu u^\mu) + \mathbb{A}(\alpha^\mu u^\nu + \alpha^\nu u^\mu) \nonumber\\
    &+& G^\alpha (l^\mu \alpha^\nu + l^\nu \alpha^\mu) + \mathbb{W}(w^\mu u^\nu + w^\nu u^\mu)\nn\\
    &+& A^w (\alpha^\mu w^\nu + \alpha^\nu w^\mu) + G^w (l^\mu w^\nu + l^\nu w^\mu)\,,
    \label{eq:general_Belinfante_Tmunu}
\end{eqnarray}
where the coefficients read:
\begin{eqnarray}
    \bar \rho &=& T^4 \frac{7\pi^4 + 30 \pi^2 \zeta^2 + 15 \zeta^4}{60 \pi^2} - 
    \frac{a^2 T^2 (\pi^2 + 3 \zeta^2)}{24 \pi^2 } - \frac{17 a^4}{960 \pi^2 }\nn\\
    &-&\frac{\omega^2 T^2 (\pi^2 + 3\zeta^2)}{8 \pi^2 } + \frac{23\, a^2\, \omega^2}{1440 \pi^2 } + \frac{\omega^4}{64\pi^2} + \frac{11(a \cdot \omega)^2}{720 \pi^2 } ,\nn\\
    G &=& T^4 \frac{\pi^2 + 3 \zeta^2}{18 \pi^2 } - \frac{31\, a^2 T^2}{360 \pi^2 } - \frac{13\, \omega^2 T^2}{120 \pi^2 },\\
    W&=& -\frac{a^2 T^2}{27\pi^2},\quad
    A= -\frac{\omega^2 T^2}{27\pi^2},\quad
    A^w = \frac{a\cdot \omega \,T^2}{27\pi^2},\nn\\
    \bar p &=& \frac{\bar\rho}{3},\quad
    G^l = -\frac{2 \,T^4}{27\pi^2 }, \quad \mathbb{A}= \mathbb{W}=G^\alpha=G^w=0\,,\nn
\end{eqnarray}
with $\zeta$ being the reduced chemical potential, $\zeta=\mu/T$.

%***********************************************************************************
\section{Coefficients of the energy-momentum tensor for free massive Dirac field}
\label{app:B}
%***********************************************************************************

The coefficients $\bar\rho$, $U_\alpha$, $U_w$, $\bar n$, $N_\alpha$, $N_w$, $A$, $W$, $\bar p$, 
$D_\alpha$, $D_w$, $G$, $G_V$, and $W^A$ in the equations \eqref{eq:Tmunu_DiracM}, \eqref{vectorcurrent} and
\eqref{axialcurrent} read~\cite{Buzzegoli:2017cqy}:
\begin{eqnarray}
\bar\rho &=&  \frac{2 m^4}{\pi^2}\Sigma \left[\frac{K_3(nx)}{nx}-\frac{K_2(nx)}{(nx)^{2}}\right] \cosh{(n\,\zeta)},\nonumber\\
\bar p&=& \frac{2m^4}{\pi^2}\,\Sigma\left[\frac{K_2(nx)}{(nx)^2}\right]\cosh(n\,\zeta),\nn\\
U_\alpha &=& \frac{m^4}{12\pi^2}\Sigma \left[(3 x^{-2} + n^2)K_2(nx)\right] \cosh{(n\,\zeta)},\nonumber\\
D_\alpha &=& \frac{m^4}{12\pi^2}\Sigma\left[-\frac{3K_2(nx)}{x^2} + n\frac{K_3(nx)}{x}\right]\cosh(n\,\zeta),\nn\\
A&=&W=0,\nn\\
D_w&=& \frac{m^4}{4\pi^2}\Sigma\left[\frac{K_2(nx)}{x^2} \right]\cosh(n\,\zeta) ,\nn\\
U_w &=&  \frac{m^4}{4\pi^2}\Sigma \left[-\frac{K_2(nx)}{x^{2}}+n  \frac{K_3(nx)}{x}\right] \cosh{(n\,\zeta)},\nonumber\\
  W^A &=&  \frac{m^3}{\pi^2}\Sigma \left[\frac{K_0(nx)}{x} + \frac{2 K_1(nx)}{n x^{2}}\right] \cosh{(n\,\zeta)},\nonumber\\
  \bar n &=& \frac{2 m^3}{\pi^2}\Sigma \left[\frac{K_2(nx)}{nx}\right] \sinh{(n\,\zeta)},\nonumber\\
  N_\alpha &=& \frac{m^3}{48 \pi^2}\Sigma \left[n^2 K_1(nx)+  3 n^2 K_3(nx)\right] \sinh{(n\,\zeta)},\nonumber\\
  N_w &=&  \frac{m^3}{4\pi^2}\Sigma \left[  \frac{n K_2(nx)}{x} \right] \sinh{(n\,\zeta)},\nn\\
  G&=&\frac{m^4}{12\pi^4}\Sigma\left[n\frac{K_3(nx)}{x}\right]\cosh(n\,\zeta),\nn\\
  G_V &=&  \frac{m^3}{6\pi^2}\Sigma \left[  \frac{n K_2(nx)}{x}\right] \sinh{(n\,\zeta)},
  \label{eq:coefficients_general}
\end{eqnarray}
where $\Sigma \equiv \Sigma_{n=1}^\infty \left(-1\right)^{n+1}$ and $x=m/T$.

\bibliographystyle{utphys}
\bibliography{pv_ref}
\end{document}